\documentclass[11pt, letterpaper]{article} 
\usepackage[margin=1in]{geometry}

\usepackage{color}
\usepackage{silence}
\usepackage{cite}
\usepackage{algorithmic}
\usepackage{algorithm}
\usepackage{amssymb}
\usepackage{amsmath}
\usepackage{multirow}
\usepackage{multicol}   
\usepackage[justification=centering]{caption}
\usepackage{footnote}
\makesavenoteenv{tabular}
\usepackage{graphicx}

\usepackage{caption}
\usepackage{subcaption}

\usepackage{epstopdf,float,amsfonts}
\usepackage{epsfig}

\newcommand{\qed}{\hfill \ensuremath{\square}}
\title{\Large \bf Imposing Robust Structured Control Constraint on Reinforcement Learning of Linear Quadratic Regulator}
\author{Sayak Mukherjee,
           Thanh Long Vu 
\thanks{ S. Mukherjee and T. L. Vu are with the Optimization and Control Group, Pacific Northwest National Laboratory (PNNL), Richland, WA, USA. 
Emails: (sayak.mukherjee, thanhlong.vu)@pnnl.gov.}}
\date{}
\begin{document}

\maketitle
\begin{abstract}
This paper discusses learning a  structured feedback control to obtain sufficient robustness to exogenous inputs for linear dynamic systems with unknown state matrix. The structural constraint on the controller is necessary for many cyber-physical systems, and our approach presents a design for any generic structure, paving the way for distributed learning control. The ideas from reinforcement learning (RL) in conjunction with control-theoretic sufficient stability and performance guarantees are used to develop the methodology. First, a model-based framework is formulated using dynamic programming to embed the structural constraint in the linear quadratic regulator (LQR) setting along with sufficient robustness conditions. Thereafter, we translate these conditions to a data-driven learning-based framework - robust structured reinforcement learning (RSRL) that enjoys the control-theoretic guarantees on stability and convergence. We validate our theoretical results with a simulation on a multi-agent network with $6$ agents.  
\end{abstract}


\section{Introduction}
Automatic control researchers over the last few decades heavily investigated control designs for large-scale and interconnected systems. Distributed control solutions are much more feasible and computationally cheap instead of performing full-scale centralized control designs for such scenarios. In many cyber-physical systems the communication infrastructure that helps to implement feedback controllers may have their own limitations. Many of such computational and practical bottlenecks have triggered a flourish in the research of distributed and structure control designs. Optimal control laws that can capture decentralized \cite{decentralized_review} or a more generic distributed \cite{thesis_distributed} structure has been investigated under notions of quadratic invariance (QI) \cite{QI1, QI2}, structured linear matrix inequalities (LMIs) \cite{dist_LMI1, dist_LMI2}, sparsity promoting optimal control \cite{sparsity_promoting} etc. \cite{geromel, geromel1989structural} discussed a structural feedback gain computation for discrete-time systems with known dynamics. 
\par
However, the above research works have been conducted with the assumption that the dynamic model of the system is known to the designer. In practice, accurate dynamic model of large-scale systems may not always be known, for example, the U.S. Eastern Interconnection transmission grid model consists of $70,000$ buses making tractable model-based control designs much difficult. Moreover, the dynamic system may also be coupled with unmodeled dynamics from coupled processes, parameter drift issues etc. which motivated research works toward learning control gains from system trajectory measurements. In recent times, ideas from machine learning and control theory are unified under the notions of reinforcement learning (RL) \cite{RL}. In the RL framework, the controller tried to optimize its policies based on the interaction with the system or environment to achieve higher rewards. Traditionally RL techniques  have been developed for the Markov Decision Process (MDP) \cite{Q,bertsekas,ADP1} based framework, and has since been the driving force to develop lots of advanced algorithmic \textit{sequential decision making} solutions. On a slightly different path, to incorporate much more rigorous and control theoretic guarantees in the context of dynamic system control where systems are modeled by differential equations, research works such as \cite{vrabie1, jiang1, V17, V18, sayak_cdc, mukherjee2021reduced} brought together the the good from the worlds of optimal and adaptive control along with the machine learning ingredients, sometimes under the notions of adaptive dynamic programming (ADP). More variants of data-driven control research such as  data-dependent linear matrix inequalities  \cite{de2019formulas}, predictive data-driven control \cite{coulson2019data}, analyzing sample complexity \cite{dean2019sample}, analysis on the characteristics of direct and indirect data-driven control \cite{dorfler2021bridging} to name a few, have been reported. Learning control designs have been reported for systems with partially or completely model free designs. Works such as \cite{jiang_assumption, jiang_book, sayak_robust, wang2017adaptive} present robust ADP/RL designs for dynamic system control. In this article, we concentrate on the learning control designs from the dynamic system viewpoint, and present a novel structured optimal control learning solution along with incorporating robustness for continuous-time linear systems with unknown state matrix. 
\par
Although reference \cite{marl2} present a survey on multi-agent and distributed RL solutions in the context of the MDPs, the area of distributed and structured learning control designs from the dynamic system viewpoint is still less explored. In \cite{mukherjee2021reduced}, a projection based reduced-dimensional RL variant have been proposed for singularly perturbed systems, \cite{Jiang_PS} presents a decentralized design for a class of interconnected systems, \cite{eth_distributed_learning} presents a structured design for discrete-time time-varying systems, \cite{fattahi2020efficient} presents a distributed learning for sparse linear systems in the recent times. Continuing in these lines of research, this article presents a robust structured optimal control learning methodology using ideas from ADP/RL. We have extended and connected some classical results that provide a bridge along our path to eventually formulating the novel RL solutions for the robust structured design, which is the core contribution of this article.
\par
We first consider the dynamic system without any exogenous perturbations, and formulate a model-based structural optimal control solution for continuous-time LTI systems using dynamic programming. Thereafter, we perform the robustness analysis in presence of exogenous inputs and provide guarantees with sufficient stability conditions. Subsequently, these model-based formulations are translated to a data-driven gain computation framework - RSRL
that can encapsulate the structural and the robustness constraints along with enjoying the stability, convergence, and sub-optimality guarantees. \textcolor{black}{It is worthy to note that the ``robustness'' of this structural feedback control to the exogenous inputs is showed in twofold. First, if the exogenous inputs are entirely unknown, then the closed-loop system is input-to-state stable to the exogenous inputs. Second, if the exogenous inputs are bounded and intermittently measurable, i.e., the exogenous inputs measurement is available at-least for 
some disjoint intervals, then the closed-loop system is globally asymptotically stable. This implies that the intermittently measurable exogenous inputs are fully compensated to compute the structured feedback controls.} We validate our design on a multi-agent dynamic network with $6$ agents. 

\section{Model and Problem formulation}
We consider a perturbed linear time-varying (LTV) continuous-time dynamic system as
\begin{align}\label{system}
    \dot{x} = Ax + B(u + \zeta(x,t)) ,\; x(0)=x_0,
\end{align}
where $x \in \mathbb{R}^{n}, u \in \mathbb{R}^{m}$ are the states and control inputs. The perturbation is caused due to the influence of the exogenous input $\zeta(x,t) \in  \mathbb{R}^m$. The function $\zeta(x,t)$ represents a functional coupling of the dynamic system with some extraneous processes which can influence the dynamic system. We consider the input matrix $B$ for the control and the exogenous input to be same, i.e., we use matched input ports. We, hereby, make the following set of assumptions.\\
\textbf{Assumption 1:} The dynamic state matrix $A$ is unknown, and the input matrix $B$ is assumed to be known.
\par
Thereafter, we characterize the nature of the coupling variable $\zeta(x,t)$ with the following assumption.\\
\textbf{Assumption 2:} The exogenous input measurements are available at-least for 
some disjoint intervals $\bigcup_{i=0}^{\infty}[t_i,t_i + \delta t] \bigcap [0,t]$ where $t$ is the current time and $\delta t$ is a small time increment \cite{jiang_assumption}, and satisfy the boundedness property given as:
\begin{align}\label{boundedness}
    \| \zeta(x,t) \|_2 \leq \alpha \| x\|_2, \alpha > 0.
\end{align}
In our numerical example, we considered $\zeta(x,t)$ to be of structure $l(t)x(t)$ with $\| l(t)\| \leq \alpha$. With this unknown state matrix, and the presence of the bounded disturbance as given in Assumption 2, we would like to learn an optimal feedback gain $u=-Kx$. However, instead of unrestricted control gain $K \in \mathbb{R}^{m \times n}$, we impose some structure on the gain. We would like to have $K \in \mathcal{K}$, which we call \textbf{\textit{structural constraint}}, where $\mathcal{K}$ is the set of all structured controllers such that: 
\begin{align}\label{struc}
    \mathcal{K} := \{ K \in \mathbb{R}^{m \times n} \; | \; F(K) = 0\}.
\end{align}
Here $F(.)$ is the matricial function that can capture any structure in the control gain matrix. This will able to implement non-zero control communication links in the $K$ matrix.
We now make the following assumption on the control gain structure $\mathcal{K}$.\\
\textbf{Assumption 3:} The communication structure required to implement the feedback control is known to the designer, and is sparse in nature.
\par
This assumption means that the structure $\mathcal{K}$ is known. This captures the limitations in the feedback communication infrastructure. For many network physical systems, the communication infrastructure can be already existing, for example, in some peer-to-peer architecture, agents can only receive feedback from their neighbors. Another very commonly designed control structure is of block-decentralized nature where local states are used to perform feedback control. Therefore, our general constraint set will encompass all such scenarios. We also make the standard stabilizability assumption.\\
\textbf{Assumption 4:} The pair $(A,B)$ is stabilizable and $(A,Q^{1/2})$ is observable, where $Q \succ 0$ is responsible to penalize the states as described in the problem statement.
\par
We can now state the problem statement as follows.\\
\textbf{P.} \textit{For system \eqref{system} satisfying Assumptions $1, 2, 3, 4$, \textbf{learn}  robust structured control policies $u = -Kx,$ where $K$ belongs to the class of structural control $\mathcal{K}$ described by \eqref{struc}, 
so that the closed-loop system is stable and the following objective is minimized}
\begin{align}\label{Ji}
    J (x(0),u^\mathcal{K}) = \int_{0}^{\infty} (x^TQx + u^TRu) d\tau.
\end{align}
    \par

\section{Structured Reinforcement Learning and Robustness}
To develop the learning control design we will take the following route. First, we consider an unperturbed dynamic system, i.e., $\dot{x} = Ax + Bu,$ and then formulate a framework where we can impose the structural constraint. Thereafter, we will consider the perturbation caused due to $\zeta(x,t)$, and then refine the framework to add robustness guarantee. 
\subsection{Structural Constraint on the Feedback Control Gain}
We start with the unperturbed dynamic system:
\begin{align}\label{sys2}
    \dot{x} = Ax + Bu, x(0) =x_0,
\end{align}
and we want to design the control $u=-Kx$ such that $K \in \mathcal{K}$.
We will first formulate a model-based solution of the optimal control problem via a modified Algebraic Riccati Equation (ARE) in continuous-time that can solve the structured optimal control when all the state matrices are known. The scenario with structural constraint without any exogenous inputs has been presented in our recent work \cite{arxiv_sayak_long}, in this article we present the detail formulation for better readability. We now present a very important model-based theorem as follows.\\
\textbf{Theorem 1:}\textit{ For any matrix $L \in \mathbb{R}^{m \times n}$, 
    let $P \succ 0$ be the solution of the following modified Riccati equation
    \begin{align}\label{struc_ARE}
       A^TP + PA -PBR^{-1}B^TP + Q + L^TRL = 0. 
    \end{align}
    Then, the control gain 
\begin{align}
\label{Lform}
    K= K(L)  = R^{-1}B^TP - L,
\end{align} will ensure closed-loop stability of \eqref{sys2}, i.e., $A-BK \in \mathbb{RH}_{\infty}$ without any external perturbation.
 \qed\\}
\textbf{Proof:} See Appendix. 

\par
 At this point, we investigate closely the matricial structure constraint. Let $I_{\mathcal{K}}$ denotes the indicator matrix for the structured matrix set $\mathcal{K}$ where this matrix contains element $1$ whenever the corresponding element in $\mathcal{K}$ is non-zero. 
The structural constraint is simply written as:
\begin{align}\label{struc2}
    F(K) = K \circ I^c_{\mathcal{K}}=\bf{0}.
\end{align}
Here, $\circ$ denotes the element-wise/ Hadamard product, and $I^c_{\mathcal{K}}$ is the complement of $I_{\mathcal{K}}$. We, hereafter, state the following theorem on the choice of $L$ to impose structure on $K$. This follows the similar form of discrete-time condition of \cite{geromel, geromel1989structural}. \\
\textbf{Theorem 2:} Let
$
    L = F(\phi(P))$ where $\phi(P) = R^{-1}B^TP.
$ Then, the control gain $K=K(L)$ designed as in Theorem 1 
will satisfy the structural constraint $F(K) = 0.$

\noindent \textbf{Proof}: We have,
\begin{align}
    K &= \phi(P) - L,\\
    &= \phi(P) - F(\phi(P)),\\
    &= \phi(P) - \phi(P)\circ I^c_{\mathcal{K}},\\
    &= \phi(P) \circ (\mathbf{1}_{m \times n} - I^c_{\mathcal{K}} ),\\
    &= \phi(P) \circ I_{\mathcal{K}} \in \mathcal{K}.
\end{align}
This concludes the proof. \qed\\ 
The implicit assumption here is the existence of the solution of the modified ARE \eqref{struc_ARE}. Once such solution exists, we will eventually learn the structured control gains. 
The next theorem describes the sub-optimality of the feasible structured solution with respect to the unconstrained objective.\\
\textbf{Theorem 3:} The difference between the feasible optimal structured control objective value $J$ and  the optimal  unstructured objective $\bar{J}$ is bounded as:
\begin{align}
\label{eq.bound}
    \| J - \bar{J}\|  \le \frac{l}{2g} \| (x_0^T \otimes x_0^T)\|,
\end{align}
for  any  feasible control structure if $\sqrt{lg \epsilon} < l/2$. Here $g=\|BR^{-1}B^T\|_2,$ $l = \| V^{-1}\|^{-1}$, where $\mathbf{V}: \mathbb{R}^{n} \to \mathbb{R}^n$ is a operator defined as:
\begin{align}
    \mathbf{V}W = (A - BR^{-1}B^T)^TW + W(A - BR^{-1}B^T).
\end{align} \qed \\
\noindent \textbf{Proof:} Let the unstructured solution of the ARE be denoted as $\bar{P}$, then the unstructured objective value is $\bar{J} = x_0^T\bar{P}x_0$, whereas, the learned structured control will result into the objective $J = x_0^TPx_0$, therefore we have,
\begin{align}
    \| J - \bar{J}\| &= \| (x_0^T \otimes x_0^T) vec (P - \bar{P})\|  \nonumber \\
    & \leq \| (x_0^T \otimes x_0^T)\| \|(P - \bar{P})\|_F 
\end{align}
Following \cite{are_pert}[Theorem 3] we use $g$, and $l$ as defined in the Theorem 3.
With $\epsilon = \frac{\|L^TRL\|}{l}$, we will have if $\sqrt{lg \epsilon} < l/2$,
\begin{align*}
    \|(P - \bar{P})\|_F &\leq \frac{2l\epsilon}{l + \sqrt{l^2 - 4lg\epsilon}} = \frac{2l\epsilon(l - \sqrt{l^2 - 4lg\epsilon})}{ 4lg\epsilon} \\
          & = \frac{(l - \sqrt{l^2 - 4lg\epsilon})}{ 2g}
          <\frac{l}{2g}.
\end{align*}

As such, the difference between the optimal values $J$ and $\bar{J}$ is bounded by
\begin{align}
\label{eq.bound}
    \| J - \bar{J}\|  \le \frac{l}{2g} \| (x_0^T \otimes x_0^T)\|
\end{align}
for any structure of the control. We note that $g$ and $l$ are not dependent on the control structure. Therefore, the inequality \eqref{eq.bound} indicates that the difference between the optimal control value $J$ with $K \in \mathcal{K}$, and optimal unstructured control value $\bar{J}$ is linearly bounded by the  initial value  $\| (x_0^T \otimes x_0^T)\|$ for any control structure.\qed    
\subsection{Robustness}    
To this end, we have formulated a modified algebraic Riccati equation that can restrict the optimal control solutions to the structural constraint. We will now investigate the robustness guarantees and necessary modifications for the system \eqref{system} with $\zeta(x(t),t)$. \\
\textbf{Theorem 4:} For any bounded exogenous input $\zeta(t)$, the structural control $u=-Kx$ where $K \in \mathcal{K}$ computed following Theorems 1 and 2 will result into the system \eqref{system} to be input-to-state stable (ISS) with respect to $\zeta(t)$. \\
\textbf{Proof:} The closed-loop dynamics with control $u=-Kx, K \in \mathcal{K}$ is given by:
\begin{align}
    \dot{x} = (A-BK)x + B\zeta(t), x(t_0)=x_0.
\end{align}
Therefore, we have
\begin{align}
    x(t) = e^{(A-BK)(t-t_0)}x_0 + \int_{t_0}^{t}e^{(A-BK)(t-\tau)}B\zeta(\tau) d\tau
\end{align}
As $A-BK$ is stable we have, $\| e^{(A-BK)(t-t_0)}\| \leq ke^{-\lambda (t-t_0)}, k>0,\lambda > 0$. Subsequently, we can bound $\| x(t)\|$ as,
\begin{align}
    \| x(t)\| \leq ke^{-\lambda (t-t_0)}\| x(t_0)\| + \frac{k\|B\|}{\lambda} \mbox{sup}_{\tau \in [t_0,t]} \|\zeta(\tau)\|.
\end{align}
Therefore, with the global asymptotic stability of the structured control, we can conclude as long as $\| \zeta(t) \|$ is bounded, the closed-loop is ISS with respect to $\zeta(t)$. \qed 

\textcolor{black}{If the exogenous inputs $\zeta(x,t)$ are intermittently measurable as in Assumption $2$, then they can be entirely compensated by the structural feedback control to ensure the global asymptotic stability of the closed-loop system, as in the following theorem which will be necessary when developing the data-driven algorithm. Please note that although the ISS condition in Theorem 4 can be ensured when bounded $\zeta(t)$ measurements are not available, as the control learning will be dependent on the state trajectories that are perturbed by the exogenous inputs, we will require the measurements of the exogenous inputs. }\\
\textbf{Theorem 5:} With assumption $2$, Let $P \succ 0$ be the solution of the following Riccati equation
\begin{align}\label{are_robust}
    (A+ \beta I)^TP + P(A + \beta I) - PBR^{-1}B^TP + Q + L^TRL = 0, \beta >0,
\end{align}
 then the control $u = -R^{-1}B^TPx + Lx$ will ensure closed-loop stability of \eqref{system} with $Q \succeq (\dfrac{\alpha^2 \lambda^2_{max}(R)}{\lambda_{min}(R)} + 2\alpha d)I$, $d \geq \mbox{max}(\|L\|)$ and $R \succ 0,$ where $\lambda_{max}(R)$ and $\lambda_{min}(R)$ are the maximum and minimum eigenvalues of $R$. \qed \\
\textbf{Proof:} See Appendix.\\
We note that the implicit assumption here is the existence of the solution of the modified ARE \eqref{struc_ARE} where
$L = F(\phi(P))$ and $\phi(P) = R^{-1}B^TP.$ It is still an open question on the necessary and sufficient condition on the structure $F(K)$ for the existence of this solution. However, once the solution exists, we can 
 iteratively compute it and the associated control gain $K$ using the algorithm formulated in the next section.
\vspace{-.45 cm}
\section{Reinforcement Learning Algorithm}    
    \noindent The gain can be iteratively computed using the following model-based  algorithm as follows. \par  
\noindent \textbf{Theorem 6: } \textit{Let $K_0$ be such that $A-BK_0$ is Hurwitz. Then, for $ k=0,1,\dots $ \\
1. Solve for ${P}_k$(Policy Evaluation) :
\begin{align}\label{Kleinman1}
\hspace{-.3 cm} &A_{k}^T{P}_k + {P}_k A_{k} + {K}_k^TR{K}_k + Q + 2 \beta P_k= 0, A_{k} = A-B{K}_k.
\end{align}
2. Update the control gain (Policy update):
\begin{align}\label{Kleinman2}
{K}_{k+1} = R^{-1}B^T {P}_k - F(\phi(P_k)), \; \phi(P_k) = R^{-1}B^T P_k.
\end{align}
Then $A - BK$ is Hurwitz and $K_{k} \in \mathcal{K}$ and $P_{k}$ converge to structured $K \in \mathcal{K}$, and $P$ as $ k  \rightarrow \infty $.}   
\qed \par
\noindent \textbf{Proof:} The theorem is formulated by taking motivation from the Kleinman's algorithm \cite{kleinman} that has been used for unstructured feedback gain computations. Comparing with the Kleinman's algorithm, the control gain update step is now modified to impose the structure. With some mathematical manipulations it can be shown that using $K_k= B^TP_k -F_k$, where $
    F_k = B^TP_k \circ I_{\mathcal{K}}^c,
$ we have,
\begin{align}
   &(A-BK_k)^TP_k + P_k(A-BK_k) + K_k^TRK_k = \\
    & A^TP_k + P_kA -P_kBR^{-1}B^TP_k + F_k^TRF_k = -Q - 2\beta P_k. 
\end{align}
Therefore, \eqref{Kleinman1} is equivalent to \eqref{are_robust} for the $k^{th}$ iteration. As we shown the stability and convergence via dynamic programming and Lyapunov analysis in Theorem $1$, considering the equivalence of Theorem $1$ with this iterative version, the theorem can be proved.   \qed
\par
\par
Although, we have formulated an iterative solution to compute structured feedback gains, the algorithm is still model-dependent. We, hereby, start to move into a state matrix agnostic design using reinforcement learning. 
We write \eqref{system} incorporating $u = -K_kx, K_k \in \mathcal{K}$ as
\begin{align}
\dot{x} &= Ax + Bu + B\zeta = (A - BK_k)x + B(K_kx + u) + B\zeta .
\end{align} 
We \textit{explore} the system by injecting a probing signal $u = u_0$ such that the system states do not become unbounded  \cite{jiang_book}. For example, following \cite{jiang_book} one can design $u_0$ as a sum of sinusoids. Thereafter, we consider a quadratic Lyapunov function $x^TPx, P \succ 0$, and we can take the time-derivative along the state trajectories, and use Theorem $6$ to alleviate the dependency on the state matrix $A$. 
\begin{align}
&\frac{d}{dt}(x^TP_kx) = x^T(A_{k}^TP_k + P_kA_{k})x + \nonumber \\ 
& \;\;\;\;\;\;\;\;\;\;\;\;\;\;\;\;\;\;\;\;\;\;\; 2(K_kx+u_{0})^TB^TP_kx + 2(\zeta^TB^T{P}_k)x, \\ 
& = -x^T(\bar{Q}_{k} + 2\beta P_k)x +  2(K_kx+u_0)^TR(K_{i(k+1)}+F_k)x + 2(\zeta^TB^T{P}_k)x, \nonumber 
\end{align}
where, $\bar{Q}_{k} = Q + K_k^TRK_k $. Therefore, starting with an arbitrary control policy $u_0$ we have,
\begin{align}
\label{main eqn}
 &\hspace{-.28 cm} x^T_{(t+T)}P_kx_{(t+T)} - x^T_t P_k x_t  + \int_{t}^{t+T}x_i^T 2\beta{P}_{k} x_i d\tau 
 -  2\int_{t}^{t+T}((K_kx+u_{0})^T (K_{k+1}+F_k)x )d\tau  \nonumber \\ & \;\;\;\;\; = \int_{t}^{t+T}(- x^T \bar{Q}_k x + 2(\zeta^TB^T{P}_k)x)d\tau.
 \end{align}
 \normalsize
 We, thereafter, solve \eqref{main eqn} by formulating an iterative algorithm using the measurements of the state and control trajectories. The algorithm basically formulates an iterative least-squares problem to solve for the structured gain. The design will require to gather data matrices $\mathcal{D} = \{ S_{xx}, T_{xx}, T_{xu_0}, T_{x\zeta}\}$ for sufficient number of time samples (discussed shortly) where $\otimes$ denotes the Kronecker product and $a|_{t_1}^{t_2} = a(t_2)-a(t_1)$ as follows:
 \begin{align} 
& \hspace{-.3 cm} S_{xx} = \begin{bmatrix}
x \otimes x |_{t_1}^{t_1+T},& \cdots &, x \otimes x |_{t_l}^{t_l+T} 
\end{bmatrix}^T,\\
& \hspace{-.3 cm} T_{xx} = \begin{bmatrix}
\int_{t_1}^{t_1+T}(x \otimes x) d\tau ,& \cdots &, \int_{t_l}^{t_l+T} (x \otimes x) d\tau \\
\end{bmatrix} ^T,\\
& \hspace{-.3 cm} T_{xu_0} = \begin{bmatrix}
\int_{t_1}^{t_1+T}(x \otimes u_0) d\tau ,& \cdots & ,\int_{t_l}^{t_l+T} (x \otimes u_0) d\tau \\
\end{bmatrix} ^T,\\
& \hspace{-.3 cm} T_{x\zeta} = \begin{bmatrix}
\int_{t_1}^{t_1+T}(x \otimes \zeta) d\tau ,& \cdots & ,\int_{t_l}^{t_l+T} (x \otimes \zeta) d\tau \\
\end{bmatrix} ^T.
\end{align} 
\vspace{-.4 cm}
 \begin{algorithm}[]
\footnotesize
\caption{ Robust Structured Reinforcement Learning (RSRL) Control}
1. \textit{Gather sufficient data:}
\textit{Store} data ($x, \zeta$ and $u_0$) for interval $(t_1,t_2,\cdots,t_l),t_i-t_{i-1}=T_{step}$.
Then \textit{construct} the following data matrices $\mathcal{D} = \{ S_{xx}, T_{xx}, T_{xu_0}, T_{x\zeta}\}$ such that rank($T_{xx} \;\; T_{xu_0} \;\; T_{x\zeta}) = n(n+1)/2 + |\mathcal{K}|+nm$. Select $Q \succeq (\dfrac{\alpha^2 \lambda^2_{max}(R)}{\lambda_{min}(R)} + 2\alpha d)I,$ where $d \geq \mbox{max}(\| F\|)$ is estimated from the allowable implementation gains.\\

2. \textit{Controller update iteration :}
Starting with a stabilizing $K_0$, \textit{Compute} $K$ iteratively ($k=0,1,\cdots$) using the following iterative equation

\textbf{for $k=0,1,2,..$}\\

A. \textit{Solve} for $P_k,$ and  $K_{k+1}+F_k$:
\begin{align}\label{eq:update}
\hspace{-.3 cm} \underbrace{\begin{bmatrix}
S_{xx} +2\beta T_{xx} & -2T_{xx}(I_n \otimes K_k^TR)  -2T_{xu_0}(I_n \otimes R) & -2T_{x\zeta}
\end{bmatrix}}_{\Theta_k}\begin{bmatrix}
vec({P}_{k}) \\ vec({K}_{(k+1)} + F_{k}) \\ vec(B^T{P}_{k})
\end{bmatrix} =\underbrace{-T_{xx}vec(\bar{Q}_{k})}_{\Phi_k}.
\end{align}
B. \textit{Compute} $F_k = R^{-1}B^TP_k \circ I_{\mathcal{K}}^c $ using the feedback structure matrix.\\ 
C. \textit{Update} the gain $K_{k+1}$.\\
D. \textit{Terminate} the loop when $||P_k - P_{k-1}|| < \varsigma$, $\varsigma > 0$ is a small threshold.\\
\textbf{endfor}\\

3. \textit{Applying K on the system :} Finally, apply $u=-Kx, K \in \mathcal{K}$, and remove $u_0$.\\
\end{algorithm} 
\normalsize

\normalsize
 Algorithm $1$ presents the steps to compute the structured feedback gain $K \in \mathcal{K}$ without knowing the state matrix $A$.

\noindent \textbf{Remark 1:} If $A$ is Hurwitz, then the controller update iteration in \eqref{eq:update} can be started without any stabilizing initial control. Otherwise, stabilizing $K_0$ is required, as commonly encountered in the RL literature \cite{jiang_book}. This is mainly due to its equivalence with modified Kleinman's algorithm in Theorem $6$.\par 
\noindent \textbf{Remark 2:} The rank condition dictates the amount of data sample needs to be gathered. For this algorithm we need  rank($T_{xx} \;\; T_{xu_{0}} \;\; T_{x \zeta}) = n(n+1)/2 + |\mathcal{K}| + nm$, where $|\mathcal{K}|$ is the number of non-zero elements in the structured feedback control matrices. This is based on the number of unknown variables in the least squares. The number of data samples can be considered to be twice this number to guarantee convergence.
\par
The condition on $Q$ in the Theorem 5 ensures robustness of the design. Although, the designer can estimate $\mbox{max}(\| L\|)$, during the initial iterations, the numerical values may not satisfy the condition on $Q$, and therefore, the designer may need to re-tune the parameter $\beta$. \\
\textbf{Theorem 7:} \textit{Performing Algorithm $1$ using $x(t), u(t)$, and $\zeta(t)$ will recover the structured ${K} \in \mathcal{K}$, and ${P}$ corresponding to Theorem 6 for (1).}\\
\textbf{Proof:}  Performing Algorithm 1 using $x(t), u(t)$, and $\zeta(t)$ is equivalently solving the trajectory relationship \eqref{main eqn}. As \eqref{main eqn} has been constructed using Theorem 6, then any solution from Theorem 6 will satisfy the $k^{th}$ iteration of the following equation:
\begin{align}\label{lemma2}
    \Theta_{k}\begin{bmatrix}
vec({P}_{k}) \\ vec({K}_{(k+1)}) \\ vec(B^T{P}_{k})
\end{bmatrix} =\Phi_{k}.
\end{align}
Hence, any solution of Theorem 6 is also a solution of \eqref{lemma2}. When the condition rank($T_{xx} \;\; T_{xu_{0}} \;\; T_{x \zeta}) = n(n+1)/2 + |\mathcal{K}| + nm$ is satisfied, $\Theta_{k}$ will have full column rank. As such, equation \eqref{lemma2} has a unique solution $ [vec({P}_{k}), vec({K}_{(k+1)}), vec(B^T{P}_{k})]^T$, from which we have a unique solution ${P}_{ik}, {K}_{i(k+1)}.$ Since this solution is unique, it is also the solution ${P}_{k}, {K}_{(k+1)}$ of theorem 6. Considering this equivalence of the Algorithm 1 with the modified Kleinman update in Theorem 6, we can conclude that the structured ${K} \in \mathcal{K}$, and ${P}$ can be recovered using Algorithm 1. \qed \par

\section{Numerical Example}

\begin{figure}[H]
    \centering
    \includegraphics[width = .7\linewidth, trim=4 4 4 10, clip]{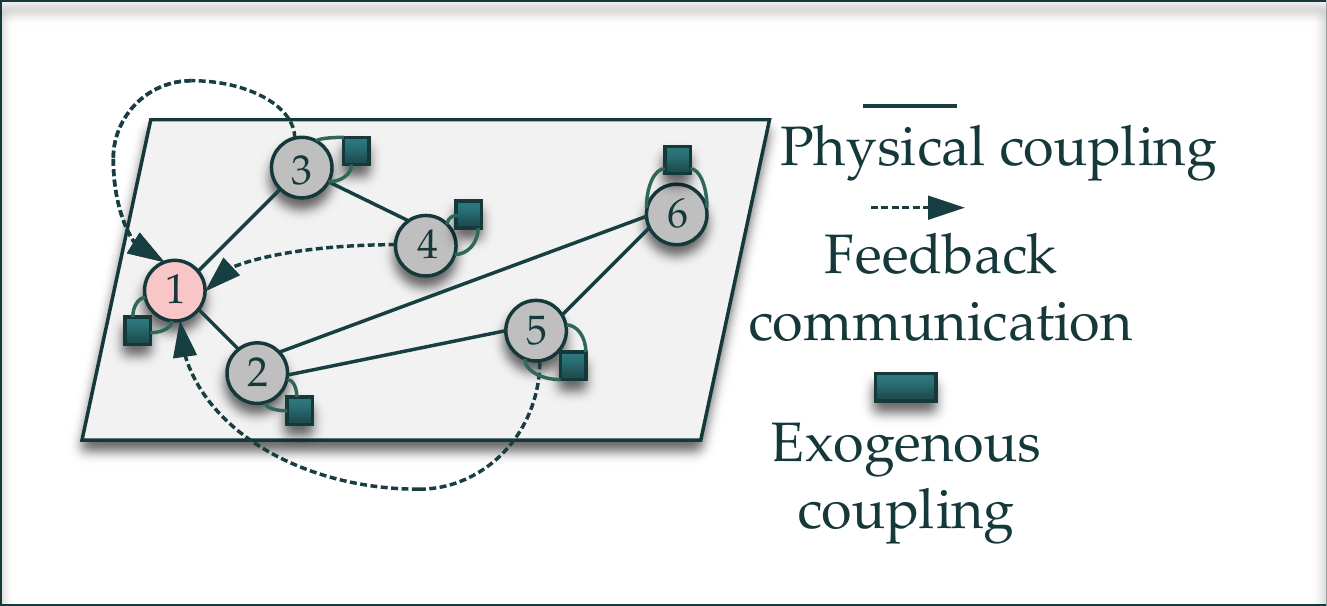}
    \caption{An example of structured feedback for agent $1$ with exogeneous inputs}
    \label{fig:scheme}
    \vspace{-.2 cm}
\end{figure}

We consider a multi-agent network with $6$ agents following the interaction structure shown in Fig. \ref{fig:scheme}. We consider each agent to follow a consensus dynamics with its neighbors such that:
\begin{align}
    \dot{x}_i = \sum_{j \in \mathcal{N}_i, i \neq j} \alpha_{ij}(x_j - x_i) + u_i +\zeta_i(x,t), x_i(0) = x_{i0}, 
\end{align}
where $\alpha_{ij} > 0$ are the coupling coefficients. We consider the state and input matrix to be:
\begin{align}
\footnotesize
    A = \begin{bmatrix} -5 &2& 3& 0& 0& 0\\
     2 &-6 &0& 0& 1& 3\\
     3 &0& -5& 2 &0 &0\\
     0 &0& 2& -2& 0& 0\\
     0 &1& 0 &0 &-4 &3\\
     0 &3& 0 &0& 3& -6 \end{bmatrix}, B=I_6.
\end{align}
\normalsize
We thereafter consider the exogeneous input perturbation to be $\zeta_i(x,t) = -0.3\cdot \mbox{sin(t)}x_i$, with the estimate of $\alpha$ to be $0.5$. We also set $\beta =1$, and $d = 2.4$ which make sure that the condition on the $Q$ can be satisfied. We consider $Q  = (3 + \alpha^2 + 2\alpha \cdot d)I_6$. The dynamics given as above is generally referred to as a Laplacian dynamics with $A.\mathbf{1_n} = \mathbf{0}$ resulting into a zero eigenvalue. We would like the controller to improve the damping of the eigenvalues closer to instability. The eigenvalues of the system are $-10.00,
   -8.27,
   -6.00,
   -3.00,
   -0.72,
   0.00$. We choose initial conditions as $[0.3,0.5,0.4,0.8,0.9,0.6]^T$. We consider an arbitrary sparsity pattern, with the assumption that the states of all the agents can be measured. We experiment with the following \textit{structure constraint}:\\ $I_\mathcal{K}(1,2) = 0, I_\mathcal{K}(1,6) = 0,  I_\mathcal{K}(2,4) = 0,I_\mathcal{K}(2,6) = 0,
I_\mathcal{K}(3,4) = 0, I_\mathcal{K}(3,5) = 0, I_\mathcal{K}(4,1) = 0, I_\mathcal{K}(4,2) = 0, I_\mathcal{K}(5,4) = 0, I_\mathcal{K}(5,3) = 0,
I_\mathcal{K}(6,4) = 0,
I_\mathcal{K}(6,1) = 0$. 

\par
\begin{figure}[t]
    \centering
        \includegraphics[width = .75\linewidth, height = 4.5 cm]{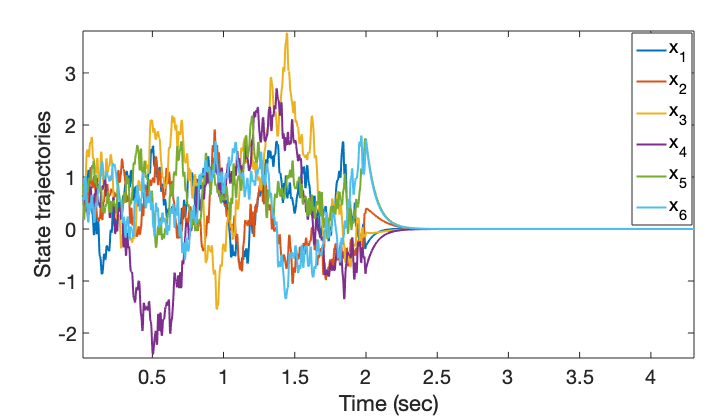} 
        \caption{State trajectories during exploration (till $2$ s) and control implementation}
        \label{fig:c1_traj}
    \begin{minipage}{0.4\linewidth}
        \centering
        \includegraphics[width = \linewidth]{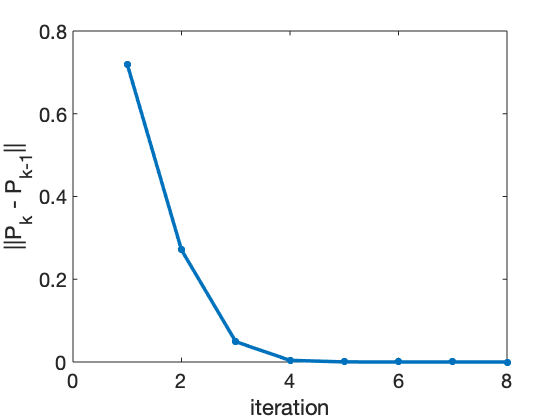} 
        \caption{ P convergence}
        \label{fig:c1p}
    \end{minipage}
    \quad
    \begin{minipage}{0.4\linewidth}
        \centering
      \includegraphics[width = \linewidth]{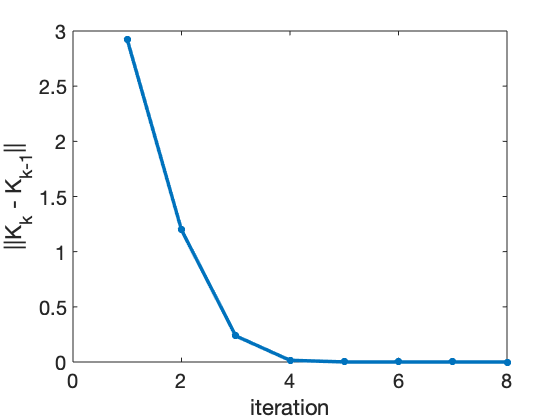} 
        \caption{K convergence}
        \label{fig:c1k}
    \end{minipage}
\vspace{-.4 cm}
\end{figure}
Here we have $n=6, m=6$, and number of non-zero elements of $K$ is $24$, therefore, we require to gather data for at-least $2(n(n+1)/2 + |\mathcal{K}| +nm)$ samples, which is $162$ data samples. We use the time step to be $0.01 s$ with an exploration of $2 s$. The iteration for $K$ and $P$ took around $0.03$ s on an average in Matlab19a with a Macbook laptop of Catalina OS, 2.8 GHz Quad-Core Intel Core i7 with 16 GB RAM. During exploration, we have used sum of sinusoids based perturbation signal to persistently excite the network. Please note that the majority of the learning is spent on the exploration of the system because of the requirement of persistent excitation and  the least square iteration is a order of magnitude smaller in comparison to the exploration time. With faster processing units, the least square iteration can be made much faster. Fig. \ref{fig:c1_traj} shows the state trajectories of the agents during exploration, and also with control implementation phase. The structured control gain learned in this scenario is given as:

\footnotesize
\begin{align}
    K_\mathcal{K}= \begin{bmatrix}
    1.7997   & 0.0000 &   0.4171&    0.0449 &   0.0170   & 0.0000\\
    0.2667 &   1.6876  &  0.0409 &  -0.0000 &   0.2021&    0.0000\\
    0.4171 &   0.0409 &   1.8017&   -0.0000  & -0.0000   & 0.0029\\
    0.0000   &-0.0000    &0.3222   & 2.2384 &  -0.0001  & -0.0001\\
    0.0170   & 0.2021  &  0.0000&    0.0000&    1.9267 &   0.4402\\
    0.0000   & 0.4038&    0.0029 &  -0.0000  &  0.4402&    1.7330
    \end{bmatrix}
\end{align}
\normalsize

The total cost comes out to be $1.0742$ units. Fig. \ref{fig:c1p}-\ref{fig:c1k} show that the $P$ and $K$ iteration converges after around $5$ iterations (using Frobenius norm). The solution also closely matches with the model-based computation.
On the other hand, the unstructured optimal control solution is given as:

\footnotesize
\begin{align}
    K_{\mbox{unstruc}} = \begin{bmatrix}
    1.6629  &  0.2407  &  0.3768    &0.0424&    0.0163&    0.0379\\
    0.2407&    1.5507  &  0.0379   & 0.0021    &0.1839   & 0.3617\\
    0.3768   & 0.0379&    1.6659&    0.2922   & 0.0009&    0.0032\\
    0.0424   & 0.0021  &  0.2922 &   2.0406  & -0.0001 &  -0.0002\\
    0.0163   & 0.1839 &   0.0009 &  -0.0001 &   1.7785 &   0.3975\\
    0.0379   & 0.3617&    0.0032  & -0.0002&    0.3975  &  1.5768\\
    \end{bmatrix},
\end{align}
\normalsize
with the optimal cost $1.068$ units. With the structured solution $K_\mathcal{K}$, the damping of the closed-loop poles is improved as they are placed at 
$-16.6912,
  -15.1893,
  -14.0494,
  -13.2785,
  -12.2264,
  -12.5005.$ This example brings out various intricacies of the algorithm and validating our theoretical results.

\section{Conclusions}
This paper presented a reinforcement learning-based robust optimal control design for linear systems with unknown state dynamics when the control is subjected to a structural constraint. We first formulate an extended algebraic Riccati equation (ARE) from the model-based analysis encompassing dynamic programming and robustness analysis with sufficient stability and convergence guarantees. Subsequently, an policy iteration based RL algorithm is formulated using the previous model-based results that continue to enjoy the rigorous guarantees and can compute the structured sub-optimal gains using the trajectory measurements of states, controls, and exogenous inputs. The sub-optimality of the learned structured gain is also quantified by comparing it with the unconstrained optimal solutions. Simulations on a multi-agent network with constrained communication infrastructure along with the exogenous influences at each agent substantiate our theoretical and algorithmic formulations. Future research will look into investigating the feasibility and methodology of robust design variants when the measurements of exogenous inputs are not available by exploiting some underlying knowledge about the exogenous input and its corresponding gain through the dynamic system.  
\bibliographystyle{plain}
\bibliography{ref}

\section{A1. Proof of Theorem 1}
We look into the optimal control solution of the unperturbed dynamic system (5) with the objective (2) using dynamic programming (DP) such that we can ensure theoretical guarantees. We assume at time $t$, the state is at $x = x_1$. We define the finite time optimal value function with the unconstrained control as:
\begin{align}
    V_t(x_1) = \mbox{min}_{u} \int_{t}^{T} (x^TQx + u^TRu) d\tau,
\end{align}
with $x(t) = x_1, \dot{x} = Ax + Bu$. Staring from state $x_1,$ the optimal $V_t(x_1)$ gives the minimum LQR cost-to-go. Now as the value function is quadratic, we can write it in a quadratic form as, $V_t(x_1)  = x_1^TP_tx_1, \; P_t \succ 0.$ We, next, look into a small time interval $[t, t+h]$, where $h >0$ is small, and in this small time interval we assume that the control is constant at $u = u_1$ and is optimal. Then cost incurred over the interval $[t, t+h]$ is 
\begin{align}
    U_1 = \int_t^{t+h} (x^TQx + u^TRu) d\tau \approx h(x_1^TQx_1 + u_1^TRu_1).
\end{align}
Also, the control input evolves the states at time $t+h$ to,
\begin{align}
    x(t+h) = x_1 + h(Ax_1 + Bu_1).
\end{align}
Then, the minimum cost-to-go from $x(t+h)$ is:
\begin{align}
    V_{t+h}(x(t+h)) &= (x_1 + h(Ax_1 + Bu_1))^T P_{t+h} (x_1 \nonumber \\  &\;\;\;\;\; +h(Ax_1 + Bu_1)).
\end{align}
Expanding $P_{t+h}$ as $(P_t + h\dot{P}_t)$ we have,
   \begin{align} 
   &V_{t+h}(x(t+h)) = \nonumber \\
    &(x_1 + h(Ax_1 + Bu_1))^T (P_t + h\dot{P}_t) (x_1 + h(Ax_1 + Bu_1)),\\
    &\approx x_1^TP_tx_1 + h((Ax_1 + Bu_1)^TP_tx_1 + x_1^TP_t(Ax_1 + Bu_1) \nonumber \\
    &+ x_1^T\dot{P}_tx_1).
\end{align}
Therefore, the total cost $U$,
\begin{align}
    V_t(x_1) &= U = U_1 + V_{t+h}(x(t+h)),\\
    &=x_1^TP_tx_1 + h(x_1^TQx_1 + u_1^TRu_1 + (Ax_1 + Bu_1)^TP_tx_1 \nonumber \\ 
    & + x_1^TP_t(Ax_1 + Bu_1) + x_1^T\dot{P}_tx_1).
\end{align}
If the control $u=u_1$ is optimal then the total cost must be minimized. Minimizing \textcolor{black}{$V_t(x_1)$} over $u_1$ we have,
\begin{align}
    u_1^TR + x_1^TP_tB = 0,\\
    u_1 = -R^{-1}B^TP_tx_1.
\end{align}
Now this gives us an optimal gain $K = R^{-1}B^TP_t$ which solves the unconstrained LQR. However, we are not interested in the unconstrained optimal gain, as that cannot impose any structure per se. In order to impose structure in the feedback gains, the feedback control will have to deviate from the optimal solution of $R^{-1}B^TP_t$, and following \cite{geromel}, we introduce another matrix $L \in \mathbb{R}^{m \times n}$ such that,
\begin{align}
    K+ L =  R^{-1}B^TP_t,\\
    K= R^{-1}B^TP_t - L.
\end{align}
The matrix $L$ will help us to impose the structure, i.e., $K \in \mathcal{K}$, which we will discuss later. Therefore, the structured implemented control is given by,
\begin{align}\label{control1}
    u_1 = -Kx_1 = -R^{-1}B^TP_tx_1 + Lx_1. 
\end{align}
We have $u_1 \in u^{\mathcal{K}}$, where $u^\mathcal{K}$ is the set of all control inputs when following $K \in \mathcal{K}$. Now, with slight abuse of notation, we denote the matrix $P_t$ to be the solution corresponding to the structured optimal control. The Hamilton-Jacobi equation with the structured control is given by,
\begin{align}
    V_t^{K \in \mathcal{K}}(x_1) &\approx \mbox{min}_{u_1 \in u^{\mathcal{K}}} U,\\
    x_1^TP_tx_1 & \approx \mbox{min}_{u_1 \in u^{\mathcal{K}}} (x_1^TP_tx_1 + h(x_1^TQx_1 + u_1^TRu_1 \nonumber \\ 
    &+ (Ax_1 + Bu_1)^TP_tx_1 + x_1^TP_t(Ax_1 + Bu_1) + \nonumber \\
    &x_1^T\dot{P}_tx_1)).
\end{align}
Putting \eqref{control1}, neglecting higher order terms, and after simplifying we get,
\begin{align}
    -\dot{P}_t = A^TP_t + P_tA -P_tBR^{-1}B^TP_t + Q + L^TRL.
\end{align}
For steady-state solution, we will have,
\begin{align}
    A^TP + PA -PBR^{-1}B^TP + Q + L^TRL  = 0. 
\end{align}
This proves the modified Riccati equation of the theorem. Now let us look into the stability of the closed-loop system with the gain $K = R^{-1}B^TP - L$. We can consider the Lyapunov function:
\begin{align}
    W = x^TPx, \; P \succ 0.
\end{align}
Therefore, the time derivative along the closed-loop trajectory of (5) is given as,
\begin{align}
    \dot{W} &= x^TP\dot{x} + \dot{x}^TPx,\\
    &= x^TP(Ax + Bu) + (Ax + Bu)^TPx,\\
    &= x^TP(Ax + B(-R^{-1}B^TPx + Lx)) + \nonumber. \\
    &\;\;\; (Ax + B(-R^{-1}B^TPx + Lx))^TPx,\\
    &= x^T[PA + A^TP - PBR^{-1}B^TP - PBR^{-1}B^TP \nonumber \\ &\;\;\;\; +PBL + L^TB^TP]x,\\
    &= x^T[-Q  - PBR^{-1}B^TP + 2PBL - L^TRL]x,\\
    &= x^T[-Q - (PBR^{-1}-L^T)R(PBR^{-1}-L^T)^T]x.
\end{align}
Now as $R$ is positive definite, the terms of form $X^TRX$ are at-least positive semi-definite. Therefore, we have,
\begin{align}
    \dot{W} \leq -x^TQx.
\end{align}
This ensures the stability of the  closed-loop system (5). Since the linear system (5) is autonomous and $(A,Q^{1/2})$ is observable, the globally asymptotic stability can be proved \textcolor{black}{by using the LaSalle's invariance principle}. This completes the proof. \qed
\section{A2. Proof of Theorem 5}
We start by considering the quadratic Lyapunov function $W = x^TPx, P \succ 0$. The closed-loop dynamics is now given by,
\begin{align}
    \dot{x} &= (A-B(R^{-1}B^TP - L)x) + B\zeta(x,t).
\end{align}
Using Assumption 2, the derivative of the Lyapunov function along the closed-loop trajectories will satisfy
\begin{align}
    \dot{W} &= x^TP(Ax + B(-R^{-1}B^TPx + Lx)) + \nonumber. \\
    &\;\;\; (Ax + B(-R^{-1}B^TPx + Lx))^TPx + 2x^TPB\zeta(x,t),\\
    &= x^T[PA + A^TP - PBR^{-1}B^TP - PBR^{-1}B^TP \nonumber \\ &\;\;\;\; +PBL + L^TB^TP]x + 2x^TPB\zeta(x,t),\\
    &= x^T[-Q  - PBR^{-1}B^TP + 2PBL - L^TRL - 2\beta P]x + 2x^TPB\zeta(x,t),\\
    &\leq x^T[-Q  - PBR^{-1}B^TP + 2PBL - L^TRL - 2\beta P]x + 2\alpha \|B^TPx \| \|x\|,\\
\end{align}

Denote $p = R^{-1}B^TPx - Lx,$ then, $B^TPx=R(p+Lx).$ Hence, $2\alpha \|B^TPx\|\|x\| = 2\alpha \|R(p+Lx)\|\|x\|\leq 2\alpha \lambda_{max}(R)(\| p\| + \|Lx\|)\| x \|$. On the other hand
\begin{align}
     p^TRp &= x^TPBR^{-1}B^TPx - x^TPBLx - x^TL^TB^TPx + x^TL^TRLx \\& 
     =x^TPBR^{-1}B^TPx - 2x^TPBLx + x^TL^TRLx. 
\end{align}
Continuing with the computation of $\dot{W}$ we have
\begin{align}
    \dot{W} &\leq x^T[-Q - 2\beta P]x - p^TRp + 2\alpha \lambda_{max}(R)\| p \|\| x \| + 2\alpha \lambda_{max}(R)\|L \| \| x \|^2,\\
    & \leq -x^T(Q -2\alpha\lambda_{max}(R)\|L\| I)x  - \lambda_{min}(R)\| p \|^2 + 2\alpha \lambda_{max}(R)\| p \|\| x \| - 2\beta \lambda_{min}(P)\| x \|^2 
\end{align}
Using $Q -2\alpha\|L\| I \succeq \dfrac{\alpha^2 \lambda^2_{max}(R)}{\lambda_{min}(R)}I$ we have,
\begin{align}
    \dot{W} &\leq -\lambda_{min}(R)(\dfrac{\alpha \lambda_{max}(R)}{\lambda_{min}(R)} \|x\| - \|p\|)^2 - 2\beta \lambda_{min}(P)\| x \|^2 \leq  - 2\beta \lambda_{min}(P)\| x \|^2  \le 0
\end{align}
Using Barbalat's lemma, it is straightforward to show that the time-varying closed-loop system (1) is globally exponentially stable at the origin. 

For implementation, as we know the structure of the feedback control, and it is practical to assume that we know the upper limits of the designable gains, we can set $d \geq \mbox{max}(\| L \|)$, and select $Q \succeq ( \dfrac{\alpha^2 \lambda^2_{max}(R)}{\lambda_{min}(R)} + 2\alpha d)I$.  
This completes the proof. \qed

\end{document}